\documentclass{optica-article}

\journal{opticajournal}
\articletype{Research Article}

\usepackage{lineno}
\usepackage{soul}
\usepackage[normalem]{ulem}
\usepackage{bm}
\usepackage{xcolor}

\begin{document}

\title{Invertible mapping between structured light and vector terahertz emission}

\author{Amirreza Sadeghpour\authormark{1}, and Daryoush Abdollahpour\authormark{1,*}}

\address{\authormark{1}Department of Physics, Institute for Advanced Studies in Basic Sciences (IASBS), Zanjan, Iran, 45137‑66731}
\email{\authormark{*}dabdollahpour@iasbs.ac.ir}

\begin{abstract*}
Terahertz (THz) radiation provides a powerful platform for ultrafast spectroscopy, imaging, and communication, yet deterministic control over its spatial and polarization structure remains challenging. Here we establish a unified framework for generating and synthesizing vectorial THz beams through coherent control of ultrafast photocurrents in semiconductors. By exploiting quantum interference between one- and two-photon excitation pathways driven by femtosecond vector beams, we demonstrate that the spatial phase and polarization structure of the optical fields can be directly mapped onto the magnitude and orientation of injected currents. This structured charge motion acts as a programmable THz antenna, enabling tailored far-field emission. Beyond forward modeling of THz generation from cylindrical vector beams and full Poincar\'e beams, we introduce an inverse-design methodology that reconstructs the required current distribution--and corresponding excitation beam profiles--from a desired THz field pattern. This invertible mapping transforms coherent photocurrent control into a systematic design strategy for THz beam shaping. Our results bridge structured light and THz photonics, providing a route toward compact, all-optical, and reconfigurable THz sources with engineered amplitude, phase, and polarization profiles.
\end{abstract*}

\section{Introduction}
\label{Intro}
In recent years, interest in both the generation and utilization of terahertz (THz) frequencies has surged. These frequencies, situated in a relatively unexplored and intriguing portion of the electromagnetic spectrum, have attracted attention due to their promising applications in areas such as spectroscopy and security imaging \cite{Fedorov18PRA, Kim09POP, Koulouklidis20NC}. A key focus has been on developing techniques to generate and detect THz radiation, which spans wavelengths from 3 mm to 30 $\mu$m \cite{Fedorov17PPCF}. Various approaches for generating THz waves have been explored, including optical rectification in nonlinear crystals \cite{Ravi15OE}, two-color filamentation \cite{Cook00OL,Kress04OL}, and photoconductive radiation from semiconductors \cite{Shan04}. 

Over the years, various techniques employing short pulses have been investigated to develop THz radiation sources \cite{Dey17NC}. A prominent approach involves generating THz radiation via photocurrent induced by two-color ultrashort pulses in gases \cite{Kim07OE,Babushkin10PRL}. In this method, a fundamental pulse (\(\omega\)) and its second-harmonic (\(2\omega\)) are combined within a gaseous medium. This interaction results in the suppression of the Coulomb barrier through rapid tunnel ionization, facilitated by the intense laser field. Due to the highly nonlinear and phase-sensitive nature of the process, ionization predominantly occurs at the peaks of the combined two-color laser field, which may differ from the peaks of the individual fields. As a consequence, electrons generated within this asymmetric laser field acquire a non-zero drift velocity, leading to the generation of directional electron currents and the simultaneous emission of THz radiation in the far-field \cite{Kim08NP}. However, for a filament exceeding the characteristic dephasing length of the two-color field, the longitudinal phase mismatch prevents forward emission. Instead, phase matching is achieved naturally through an off-axis geometry, where terahertz waves generated at different positions along the filament interfere constructively at a specific angle. This mechanism results in conical THz emission peaked at angles typically between $4^{\circ}$ and $7^{\circ}$, depending on the specific spectral frequency \cite{You12PRL}. In contrast, generating current within a semiconductor offers the potential to produce far-field THz waves with more controlled structures \cite{Kamalesh24SA}.

\begin{figure}[]
	\centering
	\includegraphics[width=.45\textwidth]{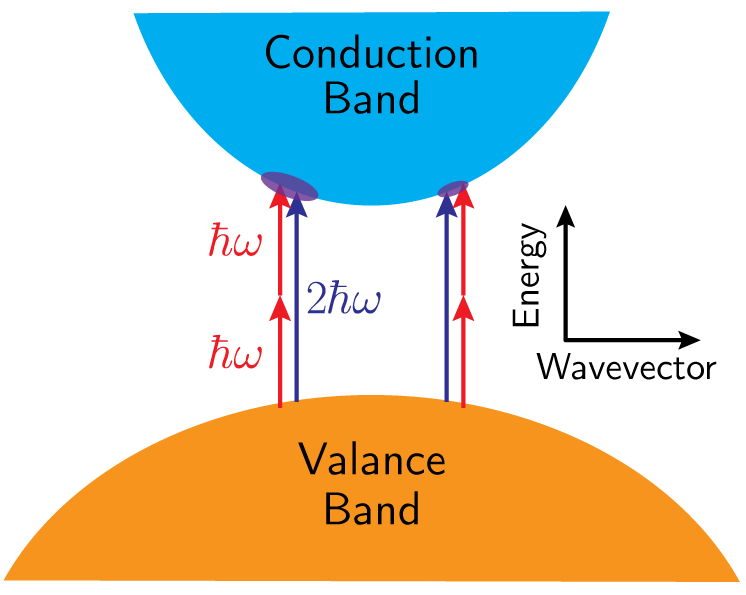}
	\caption{Quantum interference between one- and two-photon transitions in GaAs as a direct-bandgap semiconductor. Schematic band structure showing excitation pathways from the valence band to the conduction band via single-photon ($2\hbar\omega$, blue arrows) and two-photon transitions ($\hbar\omega + \hbar\omega$, red arrows). The relative phase between the two optical fields controls the direction and magnitude of the asymmetry, generating a directional current.}	
	\label{fig1}
\end{figure}

Charge currents in direct-bandgap semiconductors, such as GaAs, can be optically injected through quantum interference between distinct pathways. When the material is simultaneously illuminated by coherent optical fields at frequencies $\omega$ and $2\omega$, electrons in the valence band can reach the conduction band via two competing routes: direct single-photon absorption at $2\omega$ or two-photon absorption at $\omega$ (Fig. 1). These parallel quantum pathways connect degenerate initial and final states and therefore act as an electronic interferometer \cite{Shawn22APL}. The interference between the corresponding transition amplitudes--controlled by the relative phase--breaks the symmetry of the carrier distribution in momentum space. As a result, an asymmetric population is injected in the conduction band, giving rise to a net directional current in the absence of any external bias voltage. This process, known as quantum interference control or coherent control of photocurrents \cite{Dupont95PRL,Hache98IEEE}, enables precise tuning of both the magnitude and direction of the injected current through phase manipulation of the driving fields \cite{Shawn20PRX, Jana21NP, Kamalesh22N, Kamalesh24SA}. Previous studies demonstrated that the interference, and hence the magnitude and direction of the photocurrent, can be precisely controlled by adjusting the relative phase between the two pulses \cite{Hach97PRL, Atanasov96PRL, Shawn22APL}. These ultrafast currents are initiated on the timescale of the laser pulse duration (femtoseconds) and generate a THz electromagnetic impulse \cite{Kamalesh22N}.

Recent studies have explored the generation and control of currents in semiconductors using structured light beams. In particular, cylindrical vector beams have been investigated for their ability to excite currents and corresponding magnetic fields through both numerical simulations and experiments \cite{Shawn20NP,Jana21NP}. These studies revealed that GaAs can produce currents of up to 10 kA/cm$^2$, capable of generating magnetic fields in the range of 0.05-1 mT in the near-field \cite{Jana21NP}, with the magnetic fields numerically calculated based on experimentally measured currents. In a separate work, THz radiation amplitudes reaching the kV/cm range have been calculated from related setups \cite{Kamalesh24SA}. We identified a gap in the calculation of THz radiation in the far-field region based on current, primarily due to the limited application of near-field electric and magnetic fields. Ultrashort pulsed vector beams with spatially structured polarization profiles exhibit distinctive propagation behavior in nonlinear regimes, as recently demonstrated in air-based filamentation studies \cite{Sadeghpour25PRA}. Applying beams with such complex polarization structures to semiconductor media enables the generation of intricate current patterns and structured THz emission in the far-field. Cylindrical vector beams (CVBs) \cite{Zhan09AOP} and full Poincar\'e beams (FPBs) \cite{Beckley10OE} are two significant types of vector beams. CVBs are a subset of Poincar\'e beams characterized by axial symmetry in their polarization \cite{Zhan09AOP,Bouchard16PRL}. Full Poincar\'e beams, a specific category within Poincar\'e beams, feature polarization singularities and non-uniform polarization distributions \cite{Bouchard16PRL,Galvez12AO,Beckley10OE}, with their polarization distribution covering the entire surface of the Poincar\'e sphere.

Building upon the growing interest in polarization-resolved THz generation and control \cite{Dai09PRL,Iwase23OE,Wang24eL}, we investigate the generation of vectorial THz radiation driven by femtosecond structured light. Specifically, we focus on CVBs and FPBs, whose spatially varying polarization and phase profiles provide additional degrees of freedom for engineering ultrafast charge motion in semiconductors.
	
Our approach begins by calculating the optically injected photocurrent arising from quantum interference between one- and two-photon excitation pathways. The spatial structure of the driving vector beam imprints itself onto the current density distribution inside the semiconductor. This current is then treated as a transient antenna source, enabling the evaluation of the near-field magnetic response and the subsequent calculation of the emitted far-field THz radiation. In this forward framework, we establish a direct mapping between structured optical excitation and the resulting vectorial THz emission.

Beyond this forward modeling, we introduce an inverse-design methodology for THz beam synthesis. By reversing the conventional problem--starting from a desired far-field THz electric field distribution with specified spatial and polarization characteristics--we determine the photocurrent distribution required to generate it. From this reconstructed current, we then infer the structured optical excitation needed to realize the target THz field via coherent quantum interference control. This inverse approach transforms the problem from prediction to design, enabling programmable shaping of THz radiation through tailored femtosecond vector beams.

The manuscript is organized into two complementary parts. \textbf{Part I (“THz generation by Vector Beams”)} establishes the forward physical framework: we analyze how CVBs and FPBs drive spatially structured photocurrents and demonstrate the resulting vector THz emission characteristics. \textbf{Part II (“Inverse vector THz synthesis via quantum interference control”)} introduces and validates the inverse-design strategy, providing a systematic route to engineer far-field THz beams with prescribed amplitude, phase, and polarization distributions. Together, these two parts present both the physical foundation and the design methodology for structured THz generation based on coherent control of ultrafast currents.

\section{Part I: THz Generation by Vector Beams}

\subsection{Theoretical Model}

We begin by calculating  the two-color excitation-induced photocurrent in GaAs. As described in Introduction, the current can be controlled by inducing a two-photon transition using an electric field at an angular frequency $\omega$, together with a single-photon transition at its second-harmonic ($2\omega$). This control is achieved by tuning the relative phase between the two light waves, denoted by $\gamma_{\omega,2\omega}$ \cite{Jana21NP, Dupont95PRL}.

The two-color excitation is considered to be composed of two optical pulses at central frequencies of $\omega$ and $2\omega$, with a phase difference of $\gamma\equiv \gamma_{\omega,2\omega} $. Thus, the total applied electric field of the dual-pulse excitation can be written as a vector superposition of the electric fields of the fundamental pulse, $\mathbf{E}^{(\omega)}$, and second-harmonic pulse, $\mathbf{E}^{(2\omega)}$, as follows

\begin{equation}
	\label{eq1}
	\begin{aligned}
		\mathbf{E} (\mathbf{r},t) &= \mathbf{E}^{(\omega)} (\mathbf{r},t) + \mathbf{E}^{(2\omega)} (\mathbf{r},t) \\
		&= \mathrm{Re}\{ \bm{\mathcal{E}}^{(\omega)} (\mathbf{r},t) e^{i(kz-\omega t)} \} + \mathrm{Re}\{ \bm{\mathcal{E}}^{(2\omega)} (\mathbf{r},t) e^{i(2kz-2\omega t + \gamma)} \}
	\end{aligned}
\end{equation}
where $\bm{\mathcal{E}}^{(\omega)}$ and $\bm{\mathcal{E}}^{(2\omega)}$ are the complex-value vector envelopes of the fundamental and second harmonic pulses, respectively, and $k$ is the wavenumber of the fundamental wave. The field envelope vector can be expanded on an arbitrary orthogonal polarization basis. Here we use circular polarization basis to describe the fields with the unit vectors $\hat{\mathbf{e}}_R~=~(\hat{\mathbf{x}} - i\hat{\mathbf{y}})/\sqrt{2}$ and $\hat{\mathbf{e}}_L~=~(\hat{\mathbf{x}} + i\hat{\mathbf{y}})/\sqrt{2}$, for the  right-, and left-handed circular polarizations, respectively. The envelopes of the fundamental and second-harmonic pulses can be written as
\begin{equation}
	\label{eq2}
	\begin{aligned}
		\bm{\mathcal{E}}^{(\omega)} (\mathbf{r},t) &= \mathcal{E}^{(\omega)}_0 [\mathcal{E}^{(\omega)}_R (\mathbf{r})  \hat{\mathbf{e}}_R + \mathcal{E}^{(\omega)}_L (\mathbf{r}) \hat{\mathbf{e}}_L ] e^{\frac{-t^2}{2t^2_1}} \\
		\bm{\mathcal{E}}^{(2\omega)} (\mathbf{r},t) &= \mathcal{E}^{(2\omega)}_0 [\mathcal{E}^{(2\omega)}_R (\mathbf{r})  \hat{\mathbf{e}}_R + \mathcal{E}^{(2\omega)}_L (\mathbf{r}) \hat{\mathbf{e}}_L ] e^{\frac{-t^2}{2t^2_2}}
	\end{aligned}
\end{equation}
where $\mathcal{E}^{(\omega)}_0$ and $\mathcal{E}^{(2\omega)}_0$ denote the peak amplitudes of the fundamental and second-harmonic fields, respectively. The temporal envelopes of both pulses are assumed to be Gaussian, characterized by half-widths $t_1$ for the fundamental and $t_2$ for the second-harmonic pulses. Additionally, $\mathcal{E}_R$ and $\mathcal{E}_L$ describe the normalized spatial mode profiles of the right- and left-handed circular components. To explore the impact of polarization topology on THz generation, we model the input beam as a generalized Poincar\'e beam. The field envelope of this beam is constructed as a vector superposition of two orthogonal circular polarization states carrying independent spatial modes,
\begin{equation}
	\label{eq3}
	\bm{\mathcal{E}}(\rho,\phi,z=0) = \mathrm{LG}_{m,l}(\rho,\phi) \hat{\mathbf{e}}_L + e^{i \beta} \mathrm{LG}_{m',l'}(\rho,\phi) \hat{\mathbf{e}}_R
\end{equation}
where $\rho$ and $\phi$ are the radial and azimuthal coordinates, $\mathrm{LG}_{m,l}$ denotes the Laguerre-Gaussian mode defined by the radial index $m$ and azimuthal index $l$, and $\beta$ represents the relative phase between the right- and left-handed components. In this work, we focus on the lowest-order vector beam family; therefore, we assume both constituent modes share the same fundamental radial index ($m=m'=0$), differing only in their azimuthal indices ($l$ and $l'$). By adjusting these azimuthal indices and the phase $\beta$, a broad range of polarization morphologies can be generated, including the azimuthal and radial cylindrical vector beams (CVBs), as well as the ``star'' and ``lemon'' full Poincar\'e beams (FPBs). 

For direct-bandgap semiconductors, the evolution of the electron ($\mathbf{J}_e$) or hole ($\mathbf{J}_h$) current density during the process of quantum interference can be described by the current density injection rate equation derived by Atanasov \textit{et al.}~\cite{Atanasov96PRL}:
\begin{equation}
	\label{eq4}
	\dot{\mathbf{J}}_{e,h} = \hat{\eta}_{e,h}: \mathbf{E}^{(\omega)} \mathbf{E}^{(\omega)} \mathbf{E}^{(-2\omega)} + \text{c.c.} - \frac{\mathbf{J}_{e,h}}{\tau_{e,h}}
\end{equation}
where $\hat{\eta}_{e,h}$ is the injection current tensor, and $\tau_{e,h}$ represents the current relaxation timescale. By adjusting the phase difference between the fundamental and second-harmonic fields, a temporally asymmetric optical waveform can be generated within the semiconductor, leading to asymmetric carrier distribution in the momentum space, and hence to a directional current \cite{Hache98IEEE}. Consequently, the generated current density does not merely oscillate at optical frequencies but acquires a rectified envelope, resulting in a transient photocurrent with significant spectral components in THz range (as detailed in {\color{blue} Section 1 of Supplement}).  It is important to note that the contribution of holes to the net current is negligible compared to that of electrons due to their higher effective mass \cite{Hache98IEEE, Hach97PRL}. Consequently, $\hat{\eta}_{e} \gg \hat{\eta}_{h}$, allowing us to disregard the hole current. Therefore, we adopt this notation for the sake of simplicity: $\hat{\eta}\equiv \hat{\eta}_{e}$, and $\tau \equiv \tau_e$.

Given that the THz penetration depth in the semiconductor is on the order of 1 $\mu$m, and the THz wavelength is around 300 $\mu$m, the current along the depth of the semiconductor (represented by $z$-axis here) can be neglected \cite{Shan04}. Due to the cubic symmetry of the zinc-blende crystal structure (point group $\bar{4}3m$) of GaAs, the current injection tensor ($\hat{\eta}$) possesses only a few non-zero independent elements. Specifically, the cubic symmetry dictates that the non-zero tensor elements are invariant under coordinate permutation, leading to the relations:
\begin{equation}
	\label{eq5}
	\eta_{xxxx} = \eta_{yyyy}, \quad \eta_{xxyy} = \eta_{yyxx}, \quad \eta_{xyxy}=\eta_{yxyx} .
\end{equation}
Furthermore, for GaAs, experimental observations and theoretical models \cite{Atanasov96PRL} establish the additional relationships between these independent components:
\begin{equation}
	\label{eq6}
	\eta_{xyxy} = \eta_{xxyy} = \frac{1}{2}\eta_{xxxx}, \quad \text{and} \quad \eta_{xyyx} = 0.
\end{equation}
Therefore, the current density injection rate in $x$ direction can be expressed as  
\begin{equation}
	\label{eq7}
	\begin{aligned}
		\dot{J}_x = & (\eta_{xxxx} \mathrm{E}^{(\omega)}_x \mathrm{E}^{(\omega)}_x \mathrm{E}^{(-2\omega)}_x + \text{c.c.}) \\
			& + (\eta_{xyxy} \mathrm{E}^{(\omega)}_y \mathrm{E}^{(\omega)}_x \mathrm{E}^{(-2\omega)}_y + \text{c.c.}) \\
			& + (\eta_{xxyy} \mathrm{E}^{(\omega)}_x \mathrm{E}^{(\omega)}_y \mathrm{E}^{(-2\omega)}_y + \text{c.c.}) - \frac{J_{x}}{\tau}
	\end{aligned}
\end{equation}
where $\mathrm{E}_i^{(\omega)}$ and $\mathrm{E}_i^{(2\omega)}$ denote the $i$$^{\text{th}}$ Cartesian component ($i \in \{x, y\}$) of the complex field of the fundamental and second-harmonic pulses, respectively. A similar expression can be derived for $\dot{J}_y$, allowing for the calculation of the total current density $\mathbf{J}$. The phase difference between the fundamental pulse and its second-harmonic (i.e., $\gamma$) plays a crucial role in all terms, enabling control over both the magnitude and the direction of the current density.

Substituting the symmetry relations from Eqs. \eqref{eq5} and \eqref{eq6} into the general expression in Eq. \eqref{eq7}, we define a real-valued scalar coefficient $\eta_{0}$ such that the purely imaginary tensor component is written as $\eta_{xxxx}=i\eta_{0}$. This allows for a significant simplification of the nonlinear response. By factoring out the fundamental field component, the tensor contraction reduces to the following compact vector form
\begin{equation} 
	\label{eq8}
	\dot{\mathbf{J}} = i\eta_0 \mathbf{E}^{(\omega)} \left( \mathbf{E}^{(\omega)} \cdot \mathbf{E}^{(-2\omega)} \right) + \text{c.c.} - \frac{\mathbf{J}}{\tau}.
\end{equation}
This factorization reveals that the instantaneous direction of the generated current vector is primarily dictated by the polarization vector of the fundamental field $\mathbf{E}^{(\omega)}$, while its local magnitude and phase are modulated by the scalar inner product of the two fields, $\mathbf{E}^{(\omega)} \cdot \mathbf{E}^{(-2\omega)}$, where the overlap of the fields and their phase difference manifests. 

Therefore, the Cartesian components of the current injection rate can be written as 
\begin{equation}
	\label{eq9}
	\dot{J}_x = \frac{i \eta_0 e^{-i\gamma}}{\sqrt{2}} e^{-\left(\frac{t^2}{t^2_1} + \frac{t^2}{2t^2_2}\right)}\left( \mathcal{E}_R^{(\omega)} + \mathcal{E}_L^{(\omega)} \right) \left[ \mathcal{E}_R^{(\omega)} \mathcal{E}_R^{(-2\omega)} + \mathcal{E}_L^{(\omega)} \mathcal{E}_L^{(-2\omega)} \right] + \text{c.c.} - \frac{J_x}{\tau} ,
\end{equation}

\begin{equation}
	\label{eq10}
	\dot{J}_y = \frac{\eta_0 e^{-i\gamma}}{\sqrt{2}}e^{-\left(\frac{t^2}{t^2_1} + \frac{t^2}{2t^2_2}\right)} \left( \mathcal{E}_R^{(\omega)} - \mathcal{E}_L^{(\omega)} \right) \left[ \mathcal{E}_R^{(\omega)} \mathcal{E}_R^{(-2\omega)} + \mathcal{E}_L^{(\omega)} \mathcal{E}_L^{(-2\omega)} \right] + \text{c.c.} - \frac{J_y}{\tau} .
\end{equation}
where the factor in the square brackets represents the dot product of the fields on the circular polarization basis. These relations provide a direct mapping between the spatial phase topology of the input circular modes and the resulting Cartesian current distribution.

 While the crystal symmetry dictates the tensor relations derived above, the temporal dynamics of the generated current depend critically on both 
 the temporal widths of the input excitation pulses and the current relaxation timescale ($\tau$) dictated by the material properties. Here, we study coherent controlled current generation in low-temperature-grown GaAs (LT-GaAs). The key difference between normal GaAs and LT-GaAs lies in the carrier trapping or recombination lifetime. In standard GaAs, this lifetime is quite long, exceeding several tens of nanoseconds, whereas in LT-GaAs, the excess arsenic reduces the electron and hole lifetimes to below 1 ps \cite{Hach97PRL}. Although the nonlinear injection tensor $\hat{\eta}$ is similar for both materials, the long carrier lifetime in GaAs can result in carrier accumulation during long or high repetition rate pulsed excitation \cite{Hache98IEEE,Hach97PRL}, making LT-GaAs the preferred medium for generating the ultrafast transient currents required for THz emission.

By solving Eqs. \eqref{eq9}, and \eqref{eq10}, we obtain the current density induced within the semiconductor. As discussed, this current exhibits a transient surge-like dynamic with significant spectral content in the terahertz range. This time-varying current distribution, $\mathbf{J}(\mathbf{r},t)$, serves as the source term for the electromagnetic radiation. We proceed by calculating the retarded vector potential, $\mathbf{A}(\mathbf{r},t)$, which allows for the subsequent determination of both the electric and magnetic fields. The vector potential is related to these current density elements through the following equation \cite{Shlivinski97IEEE}
\begin{equation}
	\label{eq11}
	\mathbf{A} (\mathbf{r},t) = \frac{\mu_0}{4\pi} \frac{1}{2\pi} \int^{\infty}_{-\infty} d\omega \int d^3 r' \frac{\tilde{\mathbf{J}}(\mathbf{r}',\omega) e^{-i\omega t_r}}{|\mathbf{r} - \mathbf{r}'|}
\end{equation}
where $\mu_0$ is the vacuum permeability, and $\tilde{\mathbf{J}}(\mathbf{r}',\omega)$ represents the Fourier transform of the current density $\mathbf{J}(\mathbf{r}',t)$. Here, $\mathbf{r}'$ denotes the source coordinates located within the semiconductor plane ($z=0$), while $\mathbf{r}$ is the observation point in free space. The parameter $t_r = t - \frac{|\mathbf{r} - \mathbf{r}'|}{c}$ is the retarded time, which accounts for the propagation delay between the source and the observation point. Physically, we can treat each spatial element of the induced current density as an independent, elementary dipole. The total emitted field is the result of the coherent superposition of radiation from this array of microscopic sources. Consequently, by spatially structuring the current distribution, we effectively engineer the interference pattern of these individual dipole emissions, resulting in the generation of well-defined, sub-cycle vector THz modes that propagate through space. This coherent control mechanism offers a unique pathway towards generating ultrabroadband and reconfigurable structured THz waves \cite{Shawn22APL}.

Using the calculated vector potential $\mathbf{A}(\mathbf{r},t)$, the electromagnetic fields are determined via
\begin{align}
	\mathbf{B}(\mathbf{r},t) = \nabla \times \mathbf{A}(\mathbf{r},t), \label{eq12}\\ 
	\mathbf{E}(\mathbf{r},t) = -\frac{\partial \mathbf{A}(\mathbf{r},t)}{\partial t}, \label{eq13}
\end{align}
where we have applied the Lorentz gauge condition. The fundamental and second-harmonic pulses are modeled with central wavelengths of 1482~nm and 741~nm, and FWHM durations of 35~fs and 45~fs, respectively. We assume a beam waist of $w_0 = 200~\mu$m for both pulses. The pulse energies are set to 58 nJ for the fundamental and 1.7 nJ for the second-harmonic beam. Given our 200 $\mu$m beam waist, this yields peak intensities that are consistent with the theoretical framework established by Haché \textit{et al.}~\cite{Hache98IEEE} for bulk GaAs, ensuring the current generation remains within the efficient perturbative regime. The current relaxation timescale, $\tau$, is set to 180 femtoseconds, as reported in Ref.~\cite{Jana21NP}. In all analysis in Part I, we will consider both the fundamental and second-harmonic pulses share the same intensity, and polarization patterns. Furthermore, the value of $\eta_{0}$ is assumed to be 20 s$^{-2}$mCV$^{-3}$, based on Ref.~\cite{Atanasov96PRL}. With this theoretical framework established, we now proceed to investigate the generation of structured THz fields driven by specific vector beam topologies, starting with azimuthal and radial cylindrical vector beams.

\subsection{THz Emission from Cylindrical Vector Beams}

Different cylindrical vector beams (CVBs) can be generated by setting $m = 0$ and $l' = -l = 1$ in Eq.~\eqref{eq3}, such that the superposition of the orthogonally circular polarized LG components carries zero net OAM. The polarization morphology of the beam is determined by the phase difference ($\beta$) between the constituent components. By setting $\beta = \pi$ or $\beta = 0$, we generate azimuthally or radially polarized beams, respectively. The transverse intensity and polarization profiles of these input beams are shown in the {\color{blue} Fig. S2 of Supplement}.

As detailed in the {\color{blue} Section 2 of Supplement}, the current injection efficiency is governed by the quantum interference between the single-photon ($2\omega$) and two-photon ($\omega$) absorption pathways. This interference leads to a periodic modulation of the current magnitude dependent on the relative phase $\gamma$. For the simulations presented hereafter, we fix $\gamma$ at the optimal values that maximize the injection yield.

\begin{figure}[h!]
	\centering
	\includegraphics[width=.9\textwidth]{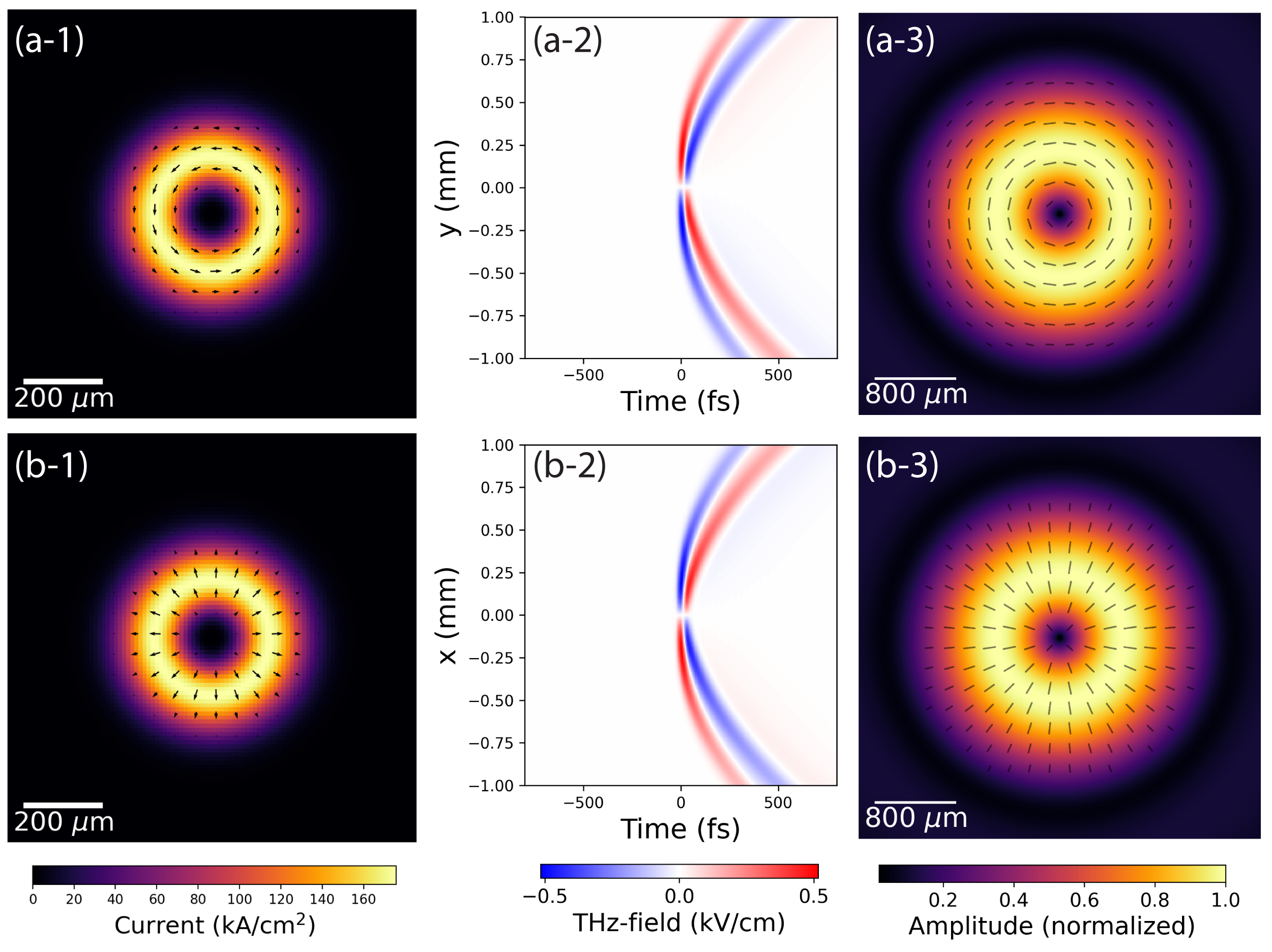}
	\caption{Current distributions and corresponding far-field THz emission characteristics. 
		(a-1), (b-1) Transverse current density profiles generated by azimuthally and radially polarized input beams, respectively. The overlaid arrows depict the local direction of the current. 
		(a-2), (b-2) Spatiotemporal evolution of the radiated far-field electric field. 
		(a-3), (b-3) Far-field amplitude distribution and polarization vector maps calculated at a frequency of 2 THz.}
	\label{Fig2}
\end{figure}

 The azimuthal and radial CVB inputs drive spatially matching azimuthal and radial current distributions, respectively, as shown in Figs. \ref{Fig2}(a-1) and \ref{Fig2}(b-1), with peak current densities reaching 175 kA/cm$^2$. To determine the resulting magnetic field ($\mathbf{B}$), we evaluate the retarded potential (Eq. \eqref{eq11}) in the near-field ($z=100\ \mu$m). Notably, the azimuthal current distribution possesses a significant spatial curl ($\nabla \times \mathbf{J}$), effectively creating a transient circulating current loop. This geometry naturally induces a strong longitudinal magnetic field along the propagation axis. Our calculations confirm that in the near-field, this azimuthal current distribution generates a magnetic field ($B_z$) exceeding 0.8 mT. The detailed spatiotemporal evolution ($x-t$) and the transverse spatial profile ($x-y$) of this magnetic field are provided in {\color{blue} Fig. S3 of Supplement}.

Beyond the near-field magnetic effects, these transient currents serve as sources for propagating THz waves. By evaluating the vector potential in the far-field ($z=4$ mm) using Eq. \eqref{eq11}, we determine the characteristics of the radiated electric field. The spatiotemporal evolution of the resulting THz pulses is illustrated in Figs. \ref{Fig2}(a-2) and \ref{Fig2}(b-2) for the azimuthal and radial CVBs, respectively. Correspondingly, Figs. \ref{Fig2}(a-3) and \ref{Fig2}(b-3) display the transverse amplitude maps and polarization vectors of the emitted fields at a frequency of $2$ THz. These results demonstrate that the far-field THz radiation faithfully preserves the vector polarization symmetry of the input optical beams. At a propagation distance of 4 mm, the peak electric field amplitude for both configurations is calculated to be 0.52 kV/cm.

\subsection{THz Emission from Full Poincar\'e Beams}
\label{Sec2-3}

\begin{figure}[]
	\centering
	\includegraphics[width=.90\textwidth]{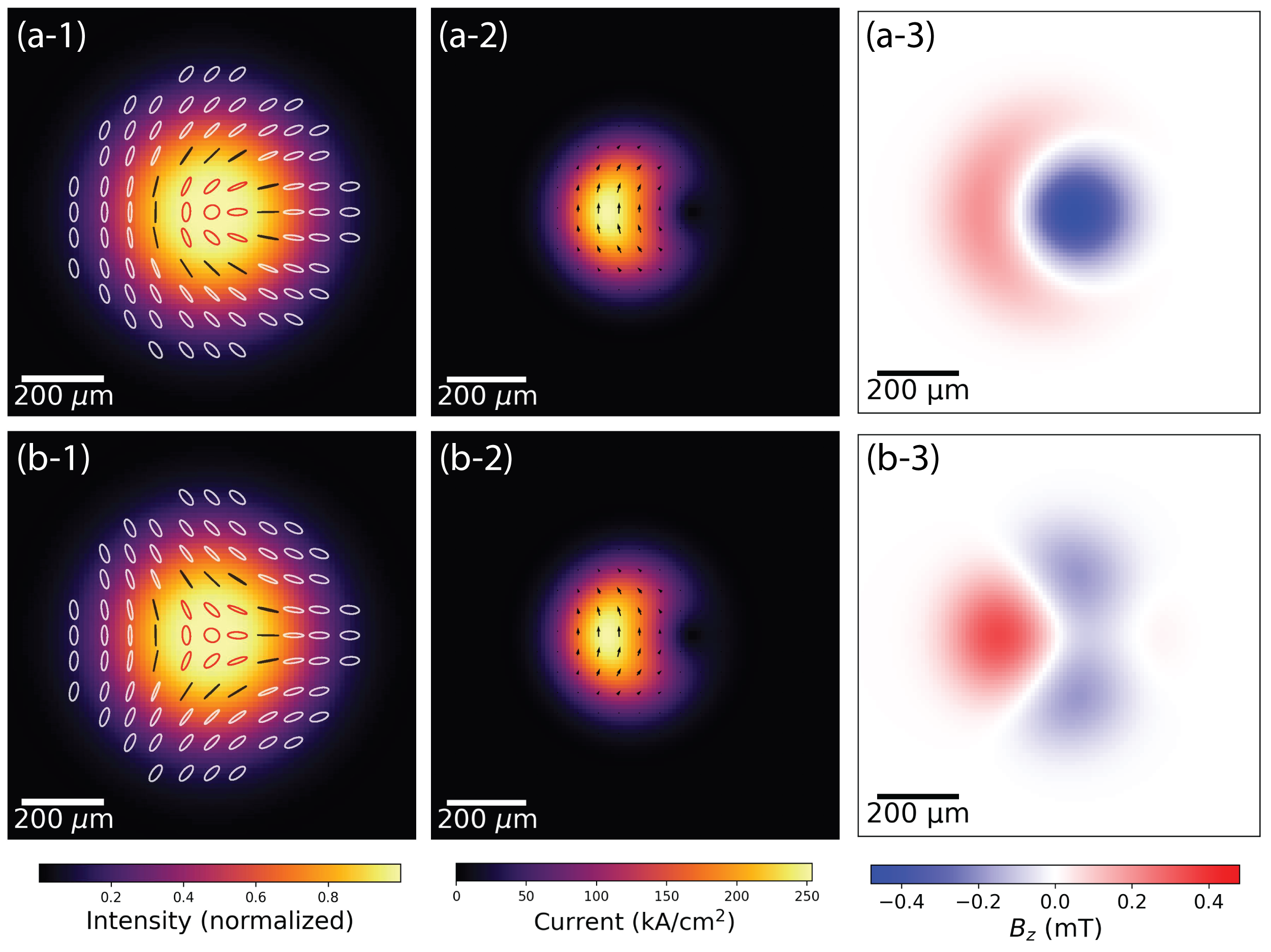}
	\caption{Current generation and near-field magnetic response driven by Full Poincar\'e Beams. The top and bottom rows correspond to the lemon and star topologies, respectively. In all cases, the fundamental and second-harmonic excitation fields share identical intensity envelopes and polarization morphologies. 
		(a-1), (b-1) Transverse intensity and polarization profiles of the input beams. The overlaid polarization ellipses depict the local state of polarization. The ellipse color indicates the polarization handedness: red for left-handed states, white for right-handed states, and black for linear polarization.
		(a-2), (b-2) Calculated current density distributions induced within the semiconductor. The overlaid arrows indicate the direction of the current.  
		(a-3), (b-3) Transverse profiles of the generated near-field longitudinal magnetic field ($B_z$).}
	\label{Fig3}
\end{figure}

While cylindrical vector beams maintain a strict cylindrical symmetry in their polarization profiles, full Poincar\'e beams (FPBs) possess a complex, non-uniform polarization topology that covers the entire surface of the Poincar\'e sphere \cite{Beckley10OE}. Different FPBs can be generated for different values of parameters $m$, $l$ and $l'$ in Eq. \eqref{eq3}. The lowest order FPBs, lemon and star FPBs, are generated by setting $l^\prime = 1$, $l^\prime = -1$, respectively, both with $m=m'=0$, $l=0$, and $\beta=0$. Figs. \ref{Fig3}(a-1) and 3(b-1) showcases the transverse intensity profiles of the lemon and star FPBs, respectively, along with their corresponding polarization topologies.

 To analyze the FPBs, we consider the case where the fundamental and second-harmonic fields share the same spatial profile defined by $\bm{\mathcal{E}} = \mathrm{LG}_{0,0} \hat{\mathbf{e}}_L + \mathrm{LG}_{0,\pm 1} \hat{\mathbf{e}}_R$. We define the real, positive radial amplitude profiles of the Gaussian and Laguerre-Gaussian modes as $A_\mathrm{G}(\rho)$ and $A_{\mathrm{LG}}(\rho)$, respectively. The complex envelopes are then given by $\mathrm{LG}_{0,0} = A_\mathrm{G}$ and $\mathrm{LG}_{0,\pm 1} = A_{\mathrm{LG}} e^{il'\phi}$. Applying the vector factorization model, the Cartesian components of the current density injection rates are initially expressed as
\begin{align}
	\dot{J}_x &\propto i(A_{\mathrm{LG}} e^{il'\phi} + A_\mathrm{G}) \left( A_\mathrm{G}^2 + A_{\mathrm{LG}}^2 \right) e^{-i\gamma} + \text{c.c.} \ - \frac{J_x}{\tau} \ , \\
	\dot{J}_y &\propto (A_{\mathrm{LG}} e^{il'\phi} - A_\mathrm{G}) \left( A_\mathrm{G}^2 + A_{\mathrm{LG}}^2 \right) e^{-i\gamma} + \text{c.c.} \ - \frac{J_y}{\tau} \ .
\end{align}
Since the scalar term $\left( A_\mathrm{G}^2 + A_{\mathrm{LG}}^2 \right)$ acts as a common positive amplitude factor and the relaxation terms $-J_{x,y}/\tau$ primarily dictate the temporal decay profile rather than the instantaneous vector direction, we neglect them to focus on the phase-dependent spatial terms. The expressions for the current components thus simplify to the explicit forms
\begin{align}
	\dot{J}_x &\propto A_\mathrm{G} \sin(\gamma) - A_{\mathrm{LG}} \sin(l'\phi - \gamma), \label{eq16} \\
	\dot{J}_y &\propto A_{\mathrm{LG}} \cos(l'\phi - \gamma) - A_\mathrm{G} \cos(\gamma).\label{eq17}
\end{align}
In this investigation, we select a control phase of $\gamma = \pi$, under which condition the components of the current density injection rate simplify to
\begin{align}
	\dot{J}_x &\propto A_{\mathrm{LG}} \sin(l'\phi), 	\label{eq18} \\
	\dot{J}_y &\propto A_\mathrm{G} - A_{\mathrm{LG}} \cos(l'\phi) .
	\label{eq19}
\end{align}
This analytical form reveals distinct current dynamics across the spatial profile of the beam. Near the beam center ($\rho < w_0/\sqrt{2}$), where the Gaussian amplitude dominates, the total current density rate is primarily directed along the positive $y$-axis. This corresponds to the strong, unidirectional current generation driven by the left-handed circular polarization of the Gaussian core, indicated by the red polarization ellipses in Figs. 3(a-1, b-1). Conversely, at large radii ($\rho > w_0/\sqrt{2}$), where the Laguerre-Gaussian mode dominates, the current follows the rotating vector field of the right-handed component, corresponding to the white polarization ellipses. Crucially, while the expression for $\dot{J}_y$ remains symmetric for both lemon ($l'=1$) and star ($l'=-1$) topologies, due to the cosine dependence, the sign of $\dot{J}_x$ flips with $l'$. This analytical result fully explains the simulation observations where the lemon and star FPBs generate currents with identical $y$-components but opposite $x$-components.

A critical interference effect occurs in the transition annulus where the Gaussian and LG amplitudes are equal ($A_\mathrm{G} \approx A_{\mathrm{LG}}$, at $\rho \approx w_0/\sqrt{2}$). In this region, the local polarization becomes linear. Substituting equal amplitudes into the simplified equations yields
\begin{align}
	\dot{J}_x &\propto \sin(l'\phi), \\
	\dot{J}_y &\propto 1 - \cos(l'\phi).
\end{align}
At the azimuthal angle $\phi=0$ and $\rho = w_0/\sqrt{2}$ (for $l'= \pm 1$), both components vanish simultaneously ($\dot{J}_x = 0$ and $\dot{J}_y = 0$). This proves analytically that destructive interference completely suppresses current generation at this specific angle and radius, creating the characteristic node or ``bean shape'' observed in the simulations. 

It is important to note that the total integrated current magnitude across the beam remains constant regardless of the value of $\gamma$. As Eqs. \eqref{eq16} and \eqref{eq17} imply, changing $\gamma$ does not alter the peak efficiency of the process but merely shifts the phase argument of the trigonometric terms. Consequently, adjusting the control phase $\gamma$ acts to azimuthally rotate the entire current distribution---including the zero-current node---without modifying the total THz yield. For higher-order FPBs ($|l'|>1$), $|l'|$ such nodes will emerge, enabling the generation of multi-lobed current structures by simply increasing the topological charge of the Laguerre-Gaussian mode.

Figures \ref{Fig3}(a-3) and \ref{Fig3}(b-3) present the transverse ($x$--$y$) distribution of the near-field longitudinal magnetic field, $B_z$. Since the magnetic field is proportional to the curl of the current density ($\mathbf{B} = \nabla \times \mathbf{A}$), its magnitude and spatial profile are critically dependent on the direction of the current vectors, which is set by the control phase $\gamma$. For the chosen phase of $\gamma = \pi$, the lemon FPB generates a peak magnetic field of approximately 0.48 mT, whereas the star FPB yields a lower peak of 0.35 mT. This disparity arises because the current distribution induced by the lemon beam exhibits a greater degree of azimuthal continuity---closely resembling a partial current loop---thereby enhancing the solenoidal magnetic field generation. In contrast, the star topology results in a more segmented current flow with reduced curl. Nevertheless, due to the non-uniform polarization gradients inherent to FPBs, both field values remain lower than that of the azimuthal CVB (0.8 mT), where the current forms a closed and symmetric loop.
\begin{figure}[]
	\centering
	\includegraphics[width=.99\textwidth]{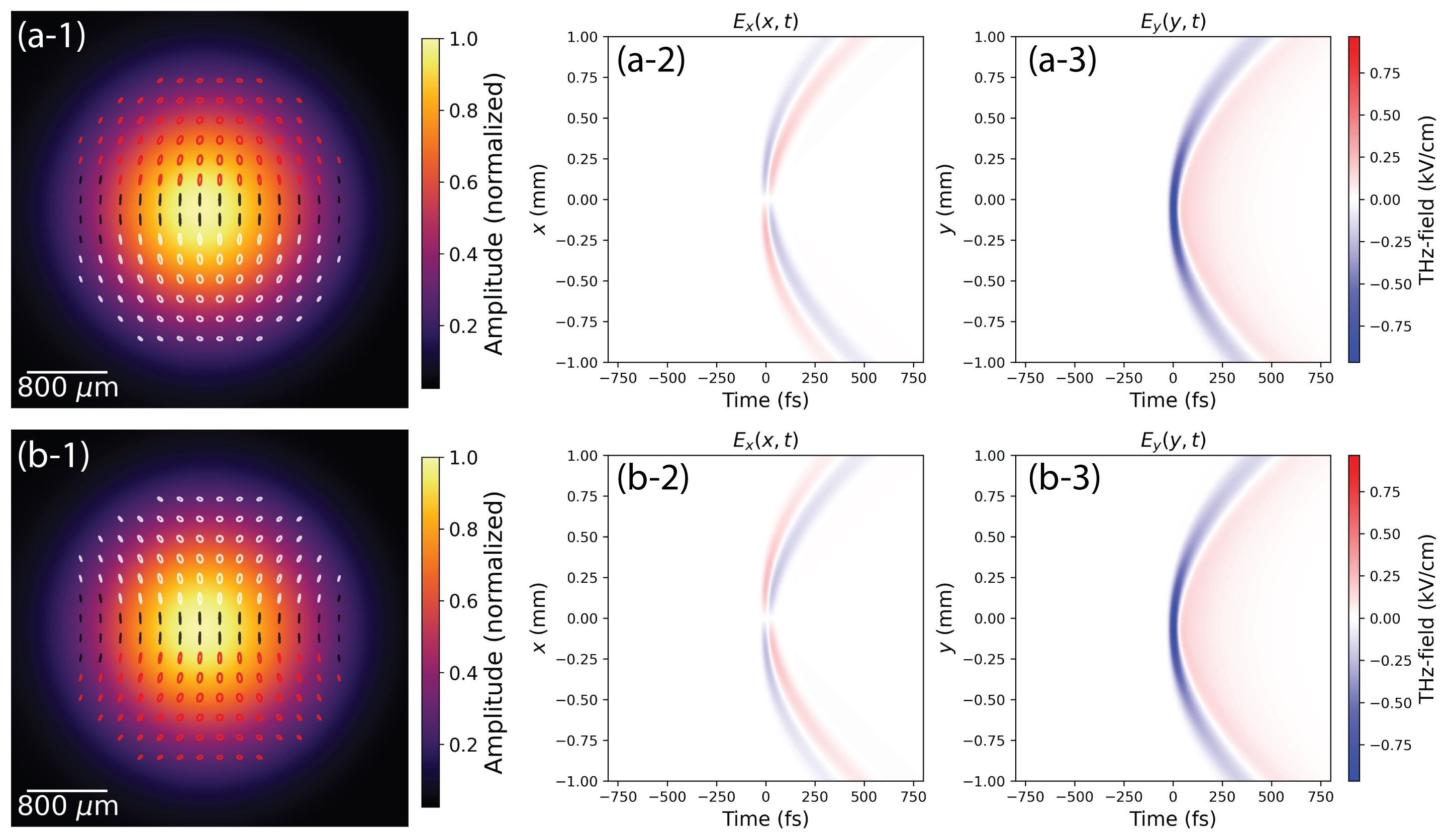}
	\caption{Far-field THz emission characteristics for Full Poincar\'e Beams (FPBs).
		(a-1), (b-1) Transverse amplitude distribution and polarization vector map of the 2 THz field generated by the lemon and star FPBs, respectively.
		(a-2), (b-2) Spatiotemporal evolution of the $x$-polarized electric field component ($E_x$) along the cut $y=0$ for the lemon and star topologies, illustrating the phase reversal between the two cases.
		(a-3), (b-3) Spatiotemporal evolution of the $y$-polarized electric field component ($E_y$) along the cut $x=0$, showing identical behavior for both beams due to the shared fundamental Gaussian mode.
		All fields are calculated at a propagation distance of $z=4$ mm.}
	\label{Fig4}
\end{figure}

 Figures \ref{Fig4}(a-1) and \ref{Fig4}(b-1) present the amplitude and polarization profiles of the far-field THz radiation at 2 THz for the lemon and star FPBs, respectively. Unlike the cylindrically symmetric cases, these beams exhibit a spatially non-uniform polarization distribution, which is a direct consequence of the complex, inhomogeneous current patterns acting as the source.

To dissect the vector nature of the emitted radiation, we analyze the spatiotemporal evolution of its orthogonal components. The far-field electric field is related to the vector potential by $\mathbf{E} = -\partial_t \mathbf{A}$, which in turn is determined by the spatial integral of the source current density (according to Eq. \eqref{eq11}). While the integration dictates the diffractive spreading and magnitude, the polarization vector of the emitted field is primarily inherited from the local direction of the source current. Thus, the symmetries observed in the components of THz emission maps ($E_x$ and $E_y$) directly reflect the analytical structure of the components of the currents ($\dot{J}_x$ and $\dot{J}_y$) derived in Eqs. \eqref{eq18} and \eqref{eq19}. Figures \ref{Fig4}(a-2, a-3) display the $E_x$ (at $y=0$) and $E_y$ (at $x=0$) components for the lemon topology, while Figs. \ref{Fig4}(b-2, b-3) show the corresponding fields for the star topology.

A distinct symmetry emerges from this comparison. As observed in the current density maps (Figs. \ref{Fig3}(a-2, b-2)), the strongest current generation occurs in the central region of the beam, where the Gaussian amplitude $A_\mathrm{G}$ dominates. Consequently, the $y$-component of the radiation is governed by the symmetric Gaussian term ($A_\mathrm{G}$) in Eq. \eqref{eq19}, which is identical for both beam topologies. This explains why the generated $E_y$ fields are virtually indistinguishable between the lemon and star cases. In contrast, the $x$-component is governed by $\dot{J}_x \propto A_{LG} \sin(l'\phi)$. Along the $y=0$ cut ($\phi=0, \pi$), this term vanishes, explaining why $E_x$ is zero on the axis. Off-axis, the sign of the field depends explicitly on the topological charge $l'$. For the lemon topology ($l'=1$), the current is positive for $0 < \phi < \pi$ (upper half-plane) and negative for $-\pi < \phi < 0$ (lower half-plane). For the star topology ($l'=-1$), this polarity is exactly reversed. This analytical dependence confirms that the observed polarity reversal in the $E_x$ radiation patterns is a direct fingerprint of the opposing orbital angular momentum of the superposed Laguerre-Gaussian modes. Furthermore, this sign flip explains the topological inversion observed in the far-field polarization maps (Figs. 4(a-1) vs 4(b-1)), while the lemon topology produces a left-handed polarization circulation in the upper lobes and right-handed in the lower lobes, the star topology exhibits the exact opposite handedness distribution.

Quantitatively, at a propagation distance of 4 mm, the peak electric field amplitude for both FPB configurations reaches about 1 kV/cm. This represents a significant enhancement compared to the CVBs analyzed in the previous section, which generated a peak field of about 0.5 kV/cm with identical excitation pulse energies. The higher yield for FPBs can be attributed to the constructive interference of the dominant Gaussian core component, which drives a strong, coherent current surge along the $y$-axis that is absent in the purely vortex-based CVB geometries.

\section{Part II: Inverse Vector THz Synthesis via Quantum Interference Control}

\subsection{Theoretical Model}
\label{sec3B}

In addition to calculating the emitted THz fields from a given optical excitation, we propose an inverse framework to determine the input conditions required to generate a desired far-field THz distribution. This approach enables the design of structured optical excitation fields (i.e., the fundamental and second-harmonic fields) and their phase difference, $\gamma(\mathbf{r})$, that yield the targeted spatial and polarization profiles of the far-field THz distribution.  

The process begins by defining a desired target electric field distribution at the far-field, $\mathbf{E}_\text{FF}$, specifying both its spatial intensity profile and polarization state, at a given frequency, in the far-field. Using vector diffraction theory, we back-propagate this target field to the source plane to determine the required near-field distribution. This reconstructed near-field profile directly dictates the necessary current density distribution within the LT-GaAs emitter. Finally, we employ quantum coherent control in the interaction of the optical excitation fields with the semiconductor to synthesize the specific required current pattern. In other words, by spatially sculpting the relative phase between the fundamental and second-harmonic excitation pulses, we can precisely imprint the target current distribution onto the semiconductor, thereby generating the desired far-field THz emission.

In order to generate a target far-field terahertz radiation with specified spatial and polarization characteristics, we first reconstruct the corresponding near-field electric field, $\mathbf{E}_\text{NF}$, from the desired far-field distribution $\mathbf{E}_\text{FF}$, using scalar diffraction theory \cite{Wang24eL},
\begin{equation}
	\label{eq22}
	\mathbf{E}_\text{NF}(\mathbf{r}) = \int \tilde{\mathbf{E}}_{\text{FF}}(\mathbf{K}_{\parallel}) G(\mathbf{K}_{\parallel}) e^{i \mathbf{K}_{\parallel}\cdot \mathbf{r}} d\mathbf{K}_{\parallel}
\end{equation}
where $\tilde{\mathbf{E}}_\text{FF}$ is the Fourier transform of $\mathbf{E}_\text{FF}$; $\mathbf{K}_{\parallel}$ denotes the transverse wavevector, and the transfer function of the free-space, $G$, is given by
\begin{equation}
	\label{eq23}
	G(\mathbf{K}_{\parallel}) = e^{iz_0 \sqrt{K_0^2 - |\mathbf{K}_{\parallel}|^2}},
\end{equation}
where $K_0$ is the wavenumber and $z_0$ is the propagation distance. Having reconstructed the required near-field distribution $\mathbf{E}_{\text{NF}}$ from the target far-field pattern, the next step is to determine the optical excitation parameters necessary to induce a source current $\mathbf{J}(\mathbf{r})$ that mimics this field profile. 
 In the Ohmic regime for the THz radiation by the induced current, the near-field THz electric field is simply proportional to the source current on the semiconductor:
 \begin{equation}
 \mathbf{J} \propto \mathbf{E}_{\text{NF}}.
 \label{eq24p}
 \end{equation}
 
To establish a direct mapping between the excitation fields and the induced current (and consequently the near-field THz electric field, $\mathbf{E}_{\text{NF}}$) we consider the fundamental and second-harmonic pulses to have circular polarizations with identical handedness. Under this condition, and in accordance with Eqs. \eqref{eq9} and \eqref{eq10}, the magnitude of the injected current is determined by the optical intensities, whereas its direction is governed exclusively by the local relative phase difference between the excitation fields, $\gamma(\mathbf{r})$. The induced current density therefore simplifies to  
   
\begin{equation}
	\label{eq24}
	\mathbf{J}(\mathbf{r}) \propto  \hat{\mathbf{x}} \sin(\gamma(\mathbf{r})) \pm \hat{\mathbf{y}} \cos(\gamma(\mathbf{r})),
\end{equation}
where, the sign of the $y$-component is determined by the handedness of the circular polarization of the excitation pulses (positive for RCP and negative for LCP). Equation~\eqref{eq24} therefore establishes a direct and deterministic correspondence between the local current orientation and the spatial phase profile of the excitation beams. In particular, the spatial variation of the relative phase, $\gamma(\mathbf{r})$ encodes the direction of the injected current at each transverse position, enabling precise control of ultrafast charge flow through phase engineering. Consequently, to synthesize a prescribed current distribution of the form $\mathbf{J}(\mathbf{r}) = J_x(\mathbf{r})\hat{\mathbf{x}} + J_y(\mathbf{r})\hat{\mathbf{y}}$ the required phase profile of the second-harmonic beam, $\gamma(\mathbf{r})$, must be tailored according to the ratio of the current components: 
\begin{equation}
	\label{eq25}
	\gamma(\mathbf{r}) = \tan^{-1}\left(\frac{J_x(\mathbf{r})}{J_y(\mathbf{r})}\right).
\end{equation}

Thus, by shaping the spatial distributions of the excitation fields to match that of the current magnitude $|\mathbf{J}(\mathbf{r})|$, and imprinting $\gamma(\mathbf{r})$ onto the wavefront of the second-harmonic pulse, one can  effectively synthesize the source current required to generate a desired far-field THz radiation pattern.

Within this framework, we assume both the fundamental and second-harmonic beams share an intensity profile proportional to the target current magnitude $|\mathbf{J}(\mathbf{r})|$. However, as indicated by Eq. \eqref{eq4}, the current generation requires the simultaneous presence of both fields. If the second-harmonic intensity vanishes in a region, no current is generated regardless of the fundamental field strength. This observation allows for a simplified experimental configuration, where one can employ a broad, collimated fundamental beam with a uniform phase and circular polarization to act as a global ``bias" field, with a proper amplitude. Under this condition, the spatial structure of the induced current is determined entirely by the second-harmonic beam, which is spatially shaped as the target amplitude $|\mathbf{J}(\mathbf{r})|$, and with the same circular polarization handedness as the fundamental beam, but with a phase difference $\gamma(\mathbf{r})$ with respect to the fundamental beam. This approach significantly reduces experimental complexity by requiring the shaping of only a single optical beam. Conversely, one could instead shape the fundamental beam while employing a collimated second-harmonic field. However, in this work, we adopt the configuration where the fundamental beam is collimated and uniform, while the spatial amplitude and phase information are encoded onto the second-harmonic beam, ensuring a direct linear mapping to the generated current distribution.

\subsection{Numerical Demonstration of THz Radiation Shaping}

To showcase the proposed inverse quantum control framework, we apply it to three representative examples that highlight its flexibility in shaping terahertz radiation. We start with two familiar profiles--a dual-Gaussian-lobe pattern and a non-diffracting Bessel beam--derived from simple target specifications. We then move to a more challenging case to demonstrate the method’s ability to reproduce complex vector structures, reconstructing the intricate THz field originally produced by the lemon FPB (Section \ref{Sec2-3}) using a completely different optical input configuration. In each case, we solve the inverse diffraction problem to determine the precise spatial phase masks needed for the second-harmonic beam, confirming that even complex THz fields can be generated using standard circularly polarized excitation.

\begin{figure}
	\centering
	\includegraphics[width=.95\textwidth]{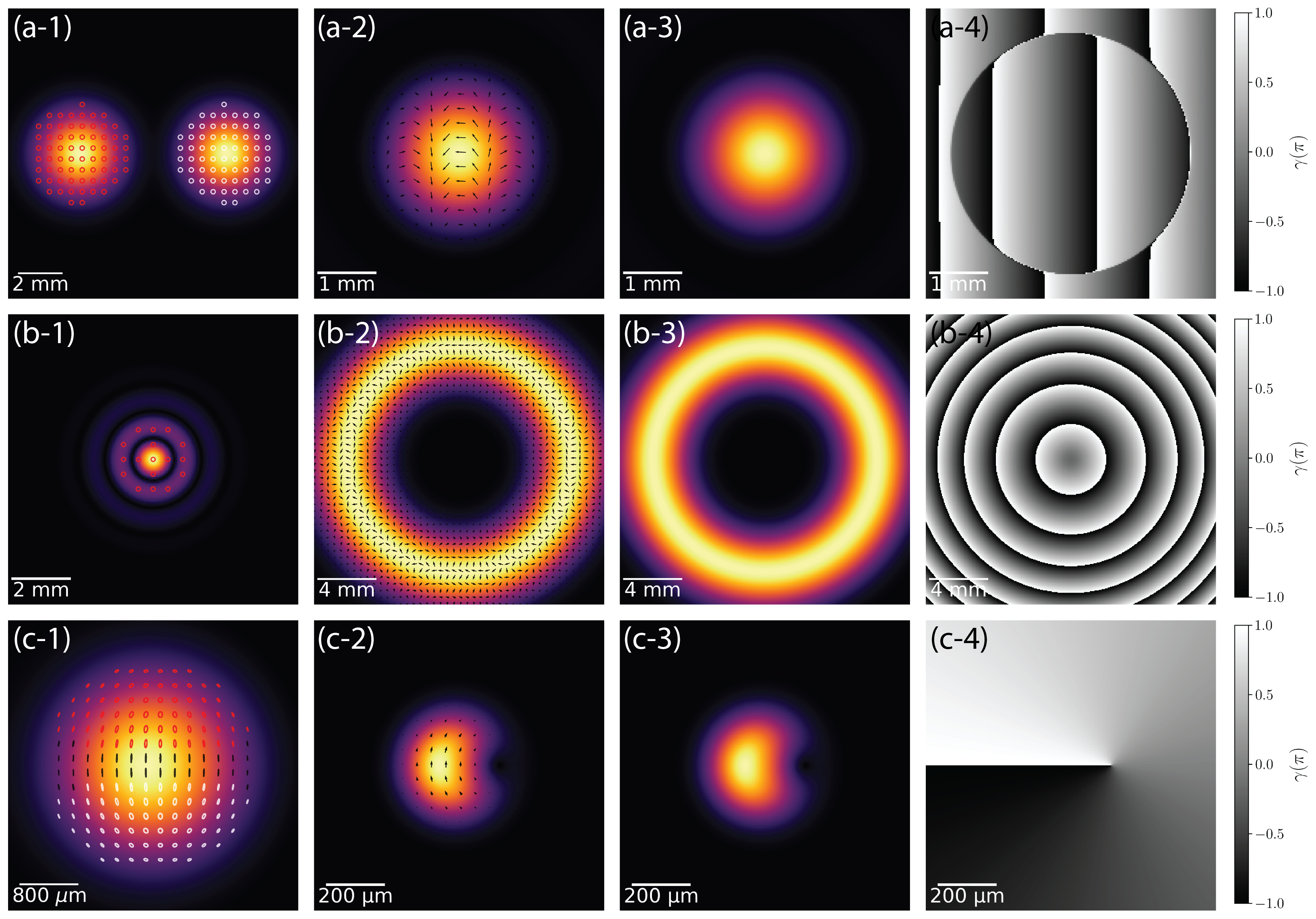}
	\caption{Inverse synthesis of structured THz radiation via coherent control.
		(a-1) Target far-field THz electric field consisting of two laterally separated Gaussian lobes with opposite circular polarizations. 
		(a-2) Reconstructed source current distribution required to generate the target field in (a-1). 
		(a-3) Amplitude profile of the left-circularly polarized second-harmonic excitation beam.
		(a-4) Spatial phase mask $\gamma(\mathbf{r})$ imposed on the second-harmonic beam to synthesize the required current through coherent control.
		(b-1)--(b-4) Inverse design of a circularly polarized zeroth-order Bessel–Gaussian THz beam: (b-1) target far-field distribution, (b-2) reconstructed current density, (b-3) amplitude of the second-harmonic beam, and (b-4) corresponding phase mask.
		(c-1)--(c-4) Reconstruction of a complex vector THz topology originally generated by a lemon full Poincar\'e beam: (c-1) target vector field with non-uniform polarization, (c-2) reconstructed current distribution, (c-3) amplitude profile of the second-harmonic excitation beam, and (c-4) phase mask encoding the required spatial phase difference. The polarization maps in the left column correspond to a frequency of 2 THz. Red and white denote left- and right-handed circular polarization, respectively, while black indicates linear polarization. In the reconstructed current density distributions (a-2), (b-2) and (c-2), the overlaid arrows depict the local direction of the current. The fundamental beam is assumed to be collimated, uniformly left-circularly polarized, and spatially homogeneous in all cases.}
	\label{Fig5}
\end{figure}

Figure \ref{Fig5}(a-1)--(a-4) illustrates the process of synthesizing a dual-Gaussian-lobe far-field THz radiation  with right-, and left-circular polarizations on each lobe at 2 THz. The target far-field THz electric field, shown in Fig \ref{Fig5}(a-1), is defined as
\begin{equation}
	\label{eq26}
	\mathbf{E}_\text{FF} = e^{-\frac{(x+x_0)^2 + y^2}{w_0^2} } e^{-ik_0x} \hat{\mathbf{e}}_L+ e^{-\frac{(x-x_0)^2 + y^2}{w_0^2}} e^{+ik_0x} \hat{\mathbf{e}}_R
\end{equation}
where $w_0$ represents the beam width, $k_0 = \frac{2\pi}{\lambda} \sin(\phi_e)$ is the transverse wavevector associated with the emission angle $\phi_e$, and $x_0 = z_0 \tan(\phi_e)$ defines the lateral displacement at the propagation distance $z_0$. For this demonstration, we select $w_0 = 1.8\,\mathrm{mm}$, $\phi_e = 30^\circ$ and $z_0 = 2\,\mathrm{cm}$. 

The reconstructed current distribution is shown in Fig.~\ref{Fig5}(a-2), while the corresponding amplitude profile of the left-circularly polarized second-harmonic beam is presented in Fig.~\ref{Fig5}(a-3). Throughout this example, the fundamental pulse is taken to be a left-circularly polarized plane wave with uniform intensity and a flat wavefront. The required spatial phase difference, $\gamma(\mathbf{r})$, between the two excitation pulses---imprinted on the second-harmonic field, for example using a spatial light modulator (SLM)---is displayed in Fig.~\ref{Fig5}(a-4).

The resulting THz profile closely resembles field structures recently demonstrated using programmable exchange-biased spintronic emitters \cite{Wang24eL}. Unlike these material-engineered approaches, however, the shaping here is achieved entirely through optical phase control of the driving field, enabling a reconfigurable and fully optical route to dynamic THz beam steering.

As a second demonstration, we apply the inverse framework to synthesize a non-diffracting THz beam in the far-field. Figure \ref{Fig5}(b-1)--(b-4) illustrate the reconstruction of a circularly polarized zeroth-order Bessel-Gaussian mode. The corresponding target far-field distribution, shown in Fig.~\ref{Fig5}(b-1), is defined as
\begin{equation}
	\label{eq27}
	\mathbf{E}_\text{FF} = J_0\left(\frac{r}{w_b}\right) e^{-(\frac{r}{w})^2} \hat{\mathbf{e}}_L,
\end{equation}
where $J_0(\cdot)$ is the zeroth-order Bessel function of the first kind, $w_b$ is proportional to the width of the central Bessel core, and $w$ represents the width of the apodizing Gaussian envelope. We define the target parameters as $w_b = 400\,\mu\mathrm{m}$ and $w = 5\,\mathrm{mm}$, calculated at a propagation distance of $z_0=5\,\mathrm{cm}$ to ensure the field has evolved into its far-field Bessel-Gaussian form.

Proceeding with the proposed backward framework, we compute the necessary source current distribution shown in Fig. \ref{Fig5}(b-2). The corresponding amplitude profile of the LCP second-harmonic excitation beam is presented in Fig. \ref{Fig5}(b-3), while the fundamental beam is considered to be a plane wave. Finally, Fig. \ref{Fig5}(b-4) displays the requisite spatial phase map $\gamma(\mathbf{r})$ that is imprinted on the second-harmonic beam.

 To further demonstrate the flexibility of the inverse coherent-control framework, we consider a more demanding test: reproducing the complex vector THz field generated earlier by a lemon full Poincar\'e beam (Section~\ref{Sec2-3}), but now using an entirely different optical excitation scheme. The target field, shown in Fig.~\ref{Fig5}(c-1), exhibits a strongly non-uniform polarization topology. In the forward approach, achieving this field required structured excitation pulses with matching lemon FPB profiles. Here, instead, we solve the inverse problem and reconstruct the same THz emission using only uniformly circularly polarized excitation beams, with the desired structure encoded solely in their spatial amplitude and relative phase.

 The reconstructed current distribution, obtained through the inverse formalism, is presented in Fig.~\ref{Fig5}(c-2). As expected, it coincides exactly with the current calculated in the forward modeling scenario (Fig.~\ref{Fig3}(a-2)), confirming the internal consistency of the approach. The corresponding amplitude profile of the second-harmonic beam and the relative phase that is considered to be imprinted on the second-harmonic beam are shown in Fig.~\ref{Fig5}(c-3), and Fig.~\ref{Fig5}(c-4), respectively. The fundamental pulse is taken to be a collimated left-circularly polarized beam with uniform intensity and flat wavefront. Notably, the optical excitation required in this inverse configuration bears no resemblance to the lemon FPB used in the forward model.

 This contrast highlights a central insight: the mapping between optical excitation and emitted THz field is inherently non-unique. Because the injected current depends on the combined amplitudes, polarization states, and relative phase of the driving fields (Eqs.~\eqref{eq9} and \eqref{eq10}), the same target current can be synthesized through qualitatively different optical configurations. One may either structure the polarization topology of the excitation beams, or employ uniform polarization and instead impose the required spatial structure through controlled amplitude and phase modulation.

 By leveraging this non-uniqueness, we show that complex vector THz fields--originally associated with structured vector-beam excitation--can be recreated using standard circularly polarized beams through spatially resolved coherent phase control. This decoupling of input polarization complexity from output vector topology establishes inverse coherent control as a powerful design principle for structured THz generation.

 The framework is readily extendable to far-field targets with increasingly complex polarization morphologies, including higher-order Poincar\'e beams, vortex beams, and other complex structured vector fields. In each case, the desired THz structure emerges from controlled modulation of the second-harmonic phase, without requiring structured polarization in the driving fields themselves. This fully optical and dynamically reconfigurable strategy establishes inverse coherent control as a general platform for engineering structured THz radiation. These results collectively demonstrate a powerful and versatile technique for designing structured THz radiation through inverse photonic design.  

\section{Conclusion}
 In this work, we have established a comprehensive framework for the generation and synthesis of vectorial terahertz radiation through coherent control of ultrafast photocurrents in semiconductors. By exploiting quantum interference between one- and two-photon excitation pathways, we demonstrated that the spatial phase and polarization structure of femtosecond vector beams can be deterministically transferred to charge motion on sub-picosecond timescales. This mapping enables structured optical fields to directly encode the magnitude and orientation of injected currents, thereby defining the spatial and polarization characteristics of the emitted THz radiation. 

Beyond forward modeling of THz generation from cylindrical vector beams and full Poincar\'e beams, we introduced an inverse-design strategy that transforms THz beam shaping into a solvable reconstruction problem. Starting from a desired far-field THz distribution, we derived the required current landscape and subsequently the structured optical excitation needed to realize it. This approach converts coherent photocurrent control from a predictive tool into a programmable design methodology. 

Our results reveal that structured light provides a powerful and versatile handle for engineering ultrafast current distributions, effectively treating the semiconductor as a reconfigurable THz antenna driven by phase-tailored femtosecond pulses. The ability to encode spatial amplitude, phase, and polarization information of THz radiation directly at the optical excitation stage opens new opportunities for compact, all-optical THz source engineering.

More broadly, this work bridges structured light, quantum interference control, and THz photonics within a unified framework. By establishing a direct and invertible correspondence between optical excitation and vector THz emission, we lay the groundwork for designer THz sources with potential applications in ultrafast spectroscopy, polarization-resolved imaging, structured THz communications, and spatiotemporal field engineering. The concepts presented here extend coherent control into the spatial domain and suggest a general route toward programmable ultrafast optoelectronic functionality.

\bibliography{references}

\end{document}